\documentclass[12pt,a4paper]{article}

\usepackage{a4wide}
\usepackage{latexsym}
\usepackage{epsf}
\usepackage{amssymb}
\usepackage{graphicx}
\usepackage{amsmath, cite}
\usepackage{amsmath,amssymb,amsthm}
\usepackage{verbatim}
\usepackage{hyperref}
\usepackage{color}

\renewcommand{\d}{\textrm{d}}
\newcommand{\e}{\textrm{e}}

\newcommand{\be}{\begin{equation}}
\newcommand{\ee}{\end{equation}}
\newcommand{\ba}{\begin{eqnarray}}
\newcommand{\ea}{\end{eqnarray}}

\renewcommand{\d}{\textrm{d}}

\begin{document}
\numberwithin{equation}{section}

\begin{center}

{\LARGE {\bf Comments on fake supersymmetry} }

\vspace{2 cm} {\large  Juan Diaz Dorronsoro, Brecht Truijen, Thomas Van Riet}\\

\vspace{1.1 cm} { Instituut voor Theoretische Fysica, KU Leuven,\\
Celestijnenlaan 200D B-3001 Leuven, Belgium}\\
\vspace{0.8 cm}

{\ttfamily {juan.diaz, brecht.truijen, thomas.vanriet @ fys.kuleuven.be }}

\vspace{3cm}
{\bf Abstract}
\end{center}

\begin{quotation}
\small Flat domain walls and spherical black holes are solutions to coupled second-order ODE's of the Hamiltonian form.  Hamilton-Jacobi theory  then implies that first-order flow equations always exist (possibly up to isolated submanifolds). If the first-order equations factorise in a specific way, they take a form that has been named fake supersymmetry. We point out that this factorisation is always possible at zero temperature. We therefore  propose a less generic definition of fake supersymmetry, which involves the boundary conditions in a non-trivial way, and we analyse its physical relevance. For instance attractor flows are necessarily fake supersymmetric in our restricted sense. To illustrate the definition we provide new analytic solutions for axion-dilaton domain walls with fake superpotentials that were argued not to exist.

\end{quotation}
\newpage
\setcounter{tocdepth}{2}
\tableofcontents 
\newpage

\section{Introduction}
The precise incarnation of holography in string theory motivated the search for solutions to supergravity theories with an asymptotic behaviour suitable for a holographic interpretation. Prime examples are domain wall solutions dual to RG flows of quantum field theories. This connection between domain walls and RG flows made it clear that the gravity solutions could (perhaps even should) obey first-order flow equations even in the absence of supersymmetry \cite{DeWolfe:1999cp, deBoer:1999xf, Skenderis:1999mm}. Indeed, as we will thoroughly discuss later on, for a zero-temperature flat domain wall consisting of a set of scalar fields $\phi^i$ spanning some Riemannian manifold with metric $h_{ij}$ plus a warpfactor $g$ from the spacetime metric, there exists a real function $\mathcal{W}(\phi)$ such that: 
\begin{align}
& \dot{\phi}^i = - h^{ij}\partial_j \mathcal{W}\,,\nonumber\\ 
& \frac{\dot{g}}{g} = \frac{\mathcal{W}}{2(D-2)} \,,\label{schematic}
\end{align}
where the dot denotes a derivative with respect to the single coordinate on which all fields depend and $D$ is the spacetime dimension. When the solution is supersymmetric, the function $\mathcal{W}$ corresponds to the genuine superpotential of the theory. Otherwise, $\mathcal{W}$ is called the \textit{fake superpotential}, and accordingly such solutions are named \textit{fake supersymmetric} (fake susy in short). Not only does the existence of this $\mathcal{W}$ make it easier to solve the equations of motion under certain conditions; fake supersymmetry also implies the non-perturbative stability of a solution \cite{Townsend:1984iu, Freedman:2003ax}. Moreover, this same structure can be found in FLRW cosmologies (through some formal Wick-rotation) \cite{Skenderis:2006jq, Skenderis:2006fb},\footnote{See \cite{Salopek:1990jq, Bazeia:2005tj} for earlier comments.} spherical black holes \cite{Ceresole:2007wx, Dall'Agata:2011nh} and general $p$-branes \cite{Janssen:2007rc}.

Early on it was realised that the existence of first-order equations for non-susy solutions (parametrised by only one coordinate) can be traced back to the Hamilton-Jacobi equations \cite{deBoer:1999xf, Townsend:2007aw, Andrianopoli:2009je}. The reason is that gravity solutions that depend on a single coordinate, like domain walls, cosmologies and spherically symmetric black holes, can be found as solutions to the effective action of a normal system in classical mechanics. Then the second-order equations can be traded for the same amount of first-order equations plus a non-linear partial differential equation called the Hamilton-Jacobi (HJ) equation. Once the HJ equation is solved, a solution to the first-order differential equations guarantees a solution to the full second-order equations. Theorems in classical mechanics imply that a solution to the HJ equation depending on the maximal amount of constants of motion always exists locally in phase space. This in turn guarantees the local existence of first-order equations.

The above suggests that fake susy is an empty concept since all solutions allow a first-order description.\footnote{Note that this is only true up to submanifolds \cite{Skenderis:2006rr}. A simple example of this subtlety are  AdS critical points (which are not flows in phase space) with tachyons below the BF bound. They cannot be fake susy since fake susy guarantees that all scalar masses are above or equal to the BF bound. But up to issues related to isolated points in phase space, it remains a fact that domain wall flows (or black hole and cosmological flows) are always fake susy in the above sense along a dense part of their trajectory.} However, this fact did not always seem to be appreciated, and it was regarded as only being valid for solutions sourced by a single scalar field. For multiple fields, on the other hand, it was stated for a while that fake susy serves as a genuine restriction in the sense that not all solutions are fake susy. Even after the single scalar result was extended in a natural way to multiple scalars \cite{Celi:2004st}, the literature on the topic remained confusing, with papers reporting on flows without a fake susy description \cite{Chemissany:2007fg, Sonner:2007cp, Perz:2008kh, Giecold:2011kf, Bobev:2013cja}.

A second look at the problem showed that things are indeed less straightforward than originally thought \cite{Trigiante:2012eb}. The equations from HJ theory are first-order gradient flows, but metric and scalars do not need to decouple as in equation (\ref{schematic}). In fact, from HJ theory one expects a general flow equation of the following style
\begin{align}
&\dot{\phi}^i =h^{ij}\partial_j\mathcal{S}(\phi, g)\,,\nonumber\\
&\dot{g} = \partial_g\mathcal{S}(\phi, g)\,,
\end{align}
with $\mathcal{S}(\phi, g)$ called Hamilton's principal function. As we review in Section \ref{folklore}, one obtains fake supersymmetric equations only when the principal function factorises as
\be\label{fact}
\mathcal{S}(\phi, g)=\mathcal{W}(\phi)F(g)\,,
\ee
with $F$ some specific function of $g$. It can be shown that this factorisation \emph{cannot} occur at non-zero temperature $T$ \cite{Perz:2008kh}, but reference \cite{Trigiante:2012eb} also presented evidence that it sometimes does not occur at zero $T$ as well. In this paper we will make the latter statement more precise, since it seems to be at odds with the arguments of \cite{Celi:2004st} which imply that the factorisation always holds at zero $T$. We will show that this apparent contradiction stems from the fact that the factorisation used in \cite{Trigiante:2012eb} was assumed to hold independently of the boundary conditions for the solutions, whereas \cite{Celi:2004st} allows superpotentials only valid for specific initial conditions for the scalars. This difference, which we make precise, allows us to define a restricted version of fake supersymmetry, based on requiring certain boundary conditions that the solutions should have. The restricted definition is well-motivated by holography and general intuition. We will also present new analytic solutions to known theories which, apart from being interesting in their own, illustrate the definition. 

The rest of the paper is organised as follows. In section \ref{folklore} we review the effective action for flat domain walls, the derivation of the fake susy equations and the link with Hamilton-Jacobi theory. In section \ref{restricted} we study in detail the factorisation property (\ref{fact}) of the principal function and we present our suggestion for a restricted notion of fake susy. Once the definition of \textit{restricted fake supersymmetry} has been introduced, we study examples in section \ref{ex}. These are newly found analytical solutions of potential interest from a holographic point of view. The physical implication are then briefly discussed in section \ref{im}.  We end with our conclusions in section \ref{discussion}. In the appendix we briefly treat black holes in section \ref{black holes}. In appendix \ref{frac} we discuss one more domain wall example, found earlier in \cite{Kuperstein:2014zda}, that does not obey the restricted fake susy condition.

\section{Effective action and fake supersymmetry folklore} \label{folklore}
In this section we recall the effective one-dimensional action for domain wall solutions in $D$ dimensions. The discussion can be directly carried over to FLRW cosmologies (see for instance \cite{Cvetic:1996vr, Skenderis:2006jq}).

Theories that support domain wall geometries are Einstein gravity coupled to $n$ scalar fields in $D$ spacetime dimensions
\begin{equation}\label{actionscalars}
S = \int\ (\star1) \left[ R - \tfrac{1}{2}h_{ij}(\phi)\partial\phi^i\partial\phi^j - V(\phi) \right]
\qquad
i, j =1,\ldots,n
\, .
\end{equation}  
The Ansatz for a domain wall solution (at finite temperature) is\footnote{We do not consider curved domain walls in this paper. Such spacetimes have a wall metric that is de Sitter or anti-de Sitter. }
\begin{equation} \label{domainwall}
ds^2_D= f(z)^2 dz^2  + g(z)^2 (-k(z)^2dt^2+d \vec{x}^2_{D-2})\,,
\end{equation}
where we allow for a blackening factor $k^2$ in front of the time component. Zero temperature implies $k^2=1$, such that we have full Poincar\'e symmetries on the wall. The function $f(z)$ in the metric is a gauge choice and can (locally) always be chosen at will. The scalar fields $\phi^i$ depend only on the coordinate $z$.

\subsection{Effective action}
If we choose a field basis in which $g$ is replaced by $ G\equiv g^{D-1}k$, the equations of motion for the different fields in the ansatz can be derived from the following effective action \cite{Kuperstein:2014zda}, for which we provide a derivation in Appendix \ref{effActionApp}
\begin{equation}\label{effAction}
I[G, k, f, \phi]=\int \mathrm{d}z\left\{
\frac{G}{f}\left[\tfrac{(D-2)}{(D-1)}\left(\frac{\dot{G}^2}{G^2}-\frac{\dot{k}^2}{k^2}\right)-\tfrac{1}{2}h_{ij}\dot{\phi}^i\dot{\phi}^j\right]
-GfV(\phi)\right\}\,.
\end{equation}
The one-dimensional nature of the effective action is at the core of the connection to Hamilton-Jacobi theory and fake supersymmetry. The $k$ e.o.m.~can be integrated as follows
\begin{equation}\label{hEOM}
\frac{G\dot{k}}{fk}= v\,,
\end{equation}
where $v$ is constant that can be interpreted as a Noether charge for the radial flow \cite{Aharony:2007vg}. Therefore $k$ can be integrated out to obtain a new effective action
\begin{equation}
I[G, f,\phi] = \int \mathrm{d}z\left\{
\frac{G}{f}\left[\tfrac{(D-2)}{(D-1)}\left(\frac{\dot{G}^2}{G^2} + \frac{f^2}{G^2}v^2\right)-\tfrac{1}{2}h_{ij}\dot{\phi}^i\dot{\phi}^j\right]
-GfV(\phi)\right\}\,.
\end{equation}
The equation of motion for $f$ is given by 
\begin{equation}\label{zeroE}
-\tfrac{(D-2)}{(D-1)}\frac{\dot{G}^2}{f^2}+\tfrac{1}{2}\frac{G^2}{f^2}h_{ij}\dot{\phi}^i\dot{\phi}^j
-G^2V(\phi) = -\tfrac{(D-2)}{(D-1)} v^2\,.
\end{equation}
For $v=0$, this has the interpretation of a zero-energy condition $\mathcal{H}=0$, with $\mathcal{H}$ the Hamiltonian of the system. When $v\neq0$, $v^2$ can be interpreted as a non-zero energy due to temperature. The constant $v$ turns out to be proportional to the product $sT$
with $s$ the entropy density and $T$ the temperature\footnote{To define the entropy density of a black domain wall we have considered domain walls that can be lifted to black branes in 10D and used the corresponding definition of entropy as derived for instance in \cite{Aharony:2007vg}.}. 
In the rest of this paper we are concerned with zero temperature since finite temperature solutions cannot be fake susy, as explained in  the comments under equation \eqref{fs3}. So from here on $v=0$ and we investigate the Hamiltonian system in $n+1$ variables $(g,\phi^i)$ and impose the zero energy constraint (\ref{zeroE}) separately.  At zero temperature and for the gauge choice $f=g^{D-1}$,  we obtain the effective action
\begin{equation}\label{effaction}
I[g,\phi] = \int \mathrm{d}z\left[  -\tfrac{1}{2}h_{ij}\dot{\phi}^i\dot{\phi}^j + (D-1)(D-2)\left( \frac{\dot{g}}{g} \right)^2   -  g^{2(D-1)} V \right]\,.
\end{equation}

\subsection{Fake supersymmetry folklore}
The Hamilton-Jacobi formulation of one-dimensional dynamical systems, such as (\ref{effaction}), is defined in terms of Hamilton's principal function $\mathcal{S}(g, \phi^i) $ that obeys the following differential equation
\be\label{HJ}
-\tfrac{1}{2}h^{ij} \partial_i\mathcal{S} \partial_j\mathcal{S} + \frac{g^2}{4(D-1)(D-2)}(\partial_g\mathcal{S})^2 + g^{2(D-1)}V = 0 \,.
\ee
We refer to this equation as the Hamilton-Jacobi (HJ) equation and to the variables $g, \phi^i$ as the generalised coordinates $q^a=(g, \phi^i)$.
Once the principal function is known, the equations of motion read
\begin{align}
\dot{\phi}^i & = -h^{ij}\partial_j\mathcal{S}\,,\nonumber\\
\dot{g} & = \frac{g^2}{2(D-1)(D-2)}\partial_g\mathcal{S}\,.\label{HJFLOW}
\end{align}
For a given function $\mathcal{S}$ that solves the HJ equation (\ref{HJ}), these are first-order differential equations for $\phi^i$ and $g$.\footnote{We should keep in mind however that it often tends to be equally difficult to solve the HJ equation (\ref{HJ}) as it is to directly solve the second-order equations of motion.}

If the principal function $\mathcal{S}$ factorises as follows \cite{Trigiante:2012eb}
\begin{equation}\label{factorisation1}
\mathcal{S} = g^{D-1}\mathcal{W}(\phi) \, , 
\end{equation} 
the Hamilton-Jacobi flow equations (\ref{HJFLOW}) become\footnote{We reinstate the function $f$ to make clear that these equations are in fact valid for all choices of $f$, not just $f=g^{D-1}$.} 
\begin{align}
& f^{-1}\dot{\phi}^i = - h^{ij}\partial_j \mathcal{W}\,,\label{fs1}\\ &
 f^{-1}\frac{\dot{g}}{g} = \frac{\mathcal{W}}{2(D-2)} \,. \label{fs2}
\end{align}
This function $\mathcal{W}$ is called a \emph{fake superpotential} \cite{Freedman:2003ax} when it is not related to supersymmetry, while for supersymmetric solutions it corresponds to the genuine \emph{superpotential}. From the HJ equation (\ref{HJ}) we see that $\mathcal{W}$ obeys the following differential equation
\begin{equation} \label{fs3}
V = \tfrac{1}{2} h^{ij}\partial_i \mathcal{W} \partial_j \mathcal{W}  - \tfrac{D-1}{4(D-2)}\mathcal{W}^2\,.
\end{equation} 
In this formulation fake  supersymmetry is equivalent to the factorisation (\ref{factorisation1}) of the principal function $\mathcal{S}$. We remark that at finite temperature a non-vanishing term containing $v^2$ would be present on the right-hand side of \eqref{HJ}, so that \eqref{factorisation1} and \eqref{fs3} would not solve the Hamilton-Jacobi equation. Note further that the factorisation implies (in the gauge $f=1$) that the scalar field equations of motion decouple from the metric since the flow equation (\ref{fs1}) only contains scalar fields and not the warpfactor $g$. This means that the solutions, describing curves in configuration space parametrized by $g$ and $\phi^i$, can be trivially projected to curves in the target space parametrized by $\phi^i$.

\subsection{Factorisable principal functions}\label{sec:fact}
The principal function ($\mathcal{S}$) is the solution to the HJ equation (\ref{HJ}) which can depend on $n+1$ arbitrary constants ($\alpha_a$). However, the functional form of $\mathcal{S}$ is not unique and hence the factorisation (\ref{factorisation1}) is ambiguous. The multiple forms of $\mathcal{S}$ originate from the various Legendre transformations one can perform on it. This is a well known fact that we recall here for completeness. 

For a given $S(q^a, \alpha_a)$ one can define
\begin{equation}
\label{eq:betas}
\beta^a \equiv \frac{\partial \mathcal{S}}{\partial \alpha_a}\,.
\end{equation} 
The Hamilton-Jacobi formalism implies that the $\beta$'s are constants of motion: $\dot{\beta}^a=0$. Starting from $\mathcal{S} \left( q^a, \alpha_a \right)$ we can always introduce a new generating function $\widetilde{\mathcal{S}} \left( q^a, \beta^a \right)$ by performing the Legendre transformation on the $\alpha$'s:
\begin{equation}
\label{Legendre}
\widetilde{\mathcal{S}} \left( q^a, \beta^a \right) = \beta^b \alpha_b - \mathcal{S}  \left( q^a, \alpha_a \right) \, .
\end{equation}
On the right-hand side, the $\alpha$'s depend on $\beta^a$ and $q^a$ through equation \eqref{eq:betas}. One easily verifies that  $\tilde{\mathcal{S}}$ solves the original HJ equation but does not provide a new solution, rather a reparametrization of the (most general) solution obtained using the original  $\mathcal{S}  \left( q^a, \alpha_a \right)$.

Given the lack of uniqueness of a principal function, one can wonder whether a Legendre transformation always exists that  brings the principal function into its factorised form.  We now argue that this is the case. 

A good hint comes from counting integration constants. Once the factorised form of $\mathcal{S}$ (\ref{factorisation1}) is assumed, the HJ equation (\ref{HJ}) becomes an equation for the (fake) superpotential (\ref{fs3}). This is a non-linear, first-order partial differential equation depending on $n$ variables $\phi^i$. We therefore expect to find $n$ integration constants inside the most general solution for $\mathcal{W}(\phi)$. This coincides with the number of integration constants in the general principal function $\mathcal{S}(g, \phi)$ since, although it depends on $n+1$ constants, one of them is a purely additive constant which can always be fixed to be zero as it does not enter the equations of motion.\footnote{In fact, this constant already vanishes implicitly in the factorised form of the principal function.}

The full proof of factorisation can be based on observations in \cite{Celi:2004st}. It proceeds by first proving that the factorisation holds in the single field case \cite{Salopek:1990jq, Bazeia:2005tj} and then relating this to the multi-field case \cite{Celi:2004st}. In what follows we take $D=4$ in order not to drag factors of $D$ along in all expressions, and choose the gauge $f=1$.

Consider a single scalar field with canonically normalised kinetic term $h_{\phi\phi}=1$. Locally, \emph{away from a critical point}, the function $\phi(z)$ can be inverted to $z(\phi)$. This can be used to consider $\dot{g}/g$ as a function of $\phi$, which we anticipatively call $\mathcal{W}$:
\be \label{Hubble}
\frac{\dot{g}}{g}(z(\phi))\equiv \tfrac{1}{4}\mathcal{W}(\phi)\,.
\ee
If we combine the second-order equation of motion for $g$ with the Hamiltonian constraint, we find
\be
\frac{\d}{\d z}\left(\frac{\dot{g}}{g}\right)=-\tfrac{1}{4}\dot{\phi}^2\,.
\ee
Together with (\ref{Hubble}), this implies:
\be
\partial\mathcal{W}\dot{\phi} = -\dot{\phi}^2\qquad\rightarrow \qquad
\dot{\phi}=-\partial \mathcal{W}(\phi)\,,
\ee
whenever $\dot{\phi}\neq 0$. Finally, if this is plugged into the Hamiltonian constraint we rediscover (\ref{fs3}) such that all equations (\ref{fs1}, \ref{fs2}, \ref{fs3}) that define a factorised principal function are obtained.

This can be extended to the multi-field case as follows \cite{Celi:2004st}. Every multi-field solution defines a curve $\mathcal{C}$ in field space $\mathcal{M}$:
\be
\mathcal{C}: \quad z\rightarrow  (\phi^1(z),\ldots \phi^n(z))\,.
\ee
For every such curve, there exists a local coordinate transformation (field redefinition) in $\mathcal{M}$, $\phi \rightarrow \tilde{\phi}$
such that all scalars but $\tilde{\phi}^1$ are zero:
\be
\mathcal{C}: \quad z\rightarrow  (\tilde{\phi}^1(z), 0, \ldots 0)\,.
\ee
This implies that the principal function, on the solution, can be taken to factorise:
\be
\mathcal{S}= g^3\mathcal{W}\Bigl(\tilde{\phi}^1[\phi^1,\ldots,\phi^n]\Bigr)\,.
\ee

It has been questioned in \cite{Sonner:2007cp} whether this reasoning is only valid when the new field redefinition is such that the truncation to the single active scalar can also be done off-shell. It is our understanding that this is not an issue since the factorisation only needs to occur for those integration constants that correspond to the solution at hand.

Note that we ignore `topological' issues with the global definition of a factorised $\mathcal{S}$-function. It is well known that $\mathcal{S}$ can be ill defined at submanifolds in phase space. For instance in the above argument for the existence of a factorisable $\mathcal{S}$ for single scalar models it was crucial that the function $\phi(z)$ was invertible to $z(\phi)$. Obviously this is in general not possible at a critical point in phase space where $\dot{\phi}=0$. This is an indication that there can be critical point solutions that are not fake supersymmetric. To construct them is straightforward: any AdS solution with tachyons below the BF bound cannot be a critical point of a fake superpotential, since a superpotential formulation restricts the scalar masses to be above the Freedman-Breithenlohner bound \cite{Skenderis:2006fb}. This argument runs in two ways since one can show that any stable AdS vacuum must allow a superpotential description \cite{Townsend:1984iu}. The local absence of a factorisable $\mathcal{S}$ is not necessarily restricted to critical points. One generically expects that there are discontinuities in $\mathcal{S}$ and its derivatives at certain points in phase space since the general $\mathcal{S}$-function along a flow might be patched together with local $\mathcal{S}$-functions and this has relevant consequences for the physics of AdS vacua, see for instance \cite{Hertog:2004ns, Amsel:2007im, Danielsson:2016rmq}.

\subsection{AdS deformed by a free scalar field }\label{Section-freeAdS}
In order to illustrate the point we have just made about factorisable principal functions, we provide here a simple explicit example. The following action
\begin{equation}
S = \int(\star1)\left[R -\tfrac{1}{2}(\partial\phi)^2 + \frac{1}{L^2}\right]\,,
\end{equation}
describes a massless field subject to a  negative cosmological constant. When the massless field is constant, the domain wall solution is simply pure AdS. When it is flowing, the asymptotic geometry is still AdS but the spacetime develops a naked singularity in the bulk \cite{Freedman:2003ax}. In the gauge $f=g^{D-1}$, the solution is a linear function 
\begin{equation}
\phi(z)= -\beta\, z + \phi(0)\,.
\end{equation}
From the fact that $\dot{\phi} = -\partial_{\phi}\mathcal{S}(\phi, g,\beta)$, we can infer that $\mathcal{S}(g,\phi,\beta)=\beta\phi+F(g,\beta)$ for some function $F$. In fact, we have
\begin{align}
\mathcal{S}(g,\phi,\beta) =& \beta \phi + \beta\sqrt{\tfrac{2(D-2)}{D-1}}\sinh^{-1}\left(\frac{\beta L}{g^{D-1}\sqrt{2}}\right)\nonumber\\
&-\frac{2g^{D-1}}{L}\sqrt{\tfrac{(D-2)}{D-1}}~\sqrt{1+\left(\frac{\beta L}{g^{D-1}\sqrt{2}}\right)^2}\,,
\end{align}
This principal function does not factorise when $\beta\neq 0$ \cite{Trigiante:2012eb}.

Through a Legendre transformation on $\beta$ we can nonetheless find a factorisable $\mathcal{S}$-function, with the following fake superpotential (found earlier in  \cite{Freedman:2003ax}) 
\begin{equation}\label{Wfreefield}
\mathcal{W}(\phi,\alpha)= \frac{2}{L}\sqrt{\tfrac{D-2}{D-1}} \, \cosh\Bigl[\sqrt{\tfrac{D-1}{2(D-2)}}(\phi -\alpha)\Bigr]\,,
\end{equation}
where $\alpha=\tfrac{\partial\mathcal{S}}{\partial\beta}$ is the Legendre transformed constant. This allows us to describe the solution as a first order flow of the fake supersymmetric form.

\section{Restricted fake supersymmetry (RFS)} \label{restricted}

Since we expect a Legendre transformation to exist that brings any $\mathcal{S}$ into the factorised form, we are led to conclude that all solutions are fake susy. In this section we will sharpen the definition of fake susy in a natural manner and get what we call \emph{restricted fake susy} (RFS). This sharper definition will follow from a well motivated physical requirement. Before presenting the definition, we  review the earlier attempt of \cite{Trigiante:2012eb} to classify solutions into fake susy and non fake susy since this forms the inspiration for our definition of RFS and gives us the opportunity to rectify some results of \cite{Trigiante:2012eb}.

\subsection{An earlier attempt}

Consider the effective action (\ref{effaction}) in the gauge $f=g^{D-1}$ and define the following quantity
\begin{equation}\label{Q}
Q = 2(D-2)\frac{\dot{g}}{g} - \mathcal{S}\,.
\end{equation}
Using the HJ equations and the second-order equation of motion for $g$, one verifies that $Q$ is a constant of motion \cite{Trigiante:2012eb}. Hence for a given domain wall solution one can compute the constant $Q$. Clearly when $\mathcal{S}$ factorises as in \eqref{factorisation1} we must have $Q=0$. But we can also run the argument the other way around, by substituting the HJ flow equation for $g$ 
\begin{equation}
\frac{\dot{g}}{g} =\frac{g\partial_g\mathcal{S}}{2(D-1)(D-2)}\,,
\end{equation}
in equation (\ref{Q}). Hence this shows that fake  supersymmetry can be neatly characterized as 
\begin{equation}\label{fake supersymmetry1}
\text{Fake supersymmetry}\Longleftrightarrow Q=0\,.
\end{equation}
So not all solutions can be fake supersymmetric at first sight and indeed some explicit examples were given in \cite{Trigiante:2012eb} that had non-zero $Q$ (the latter examples were used earlier in \cite{Sonner:2007cp, Chemissany:2007fg} to argue that not all solutions are fake  susy). However, the condition (\ref{fake supersymmetry1}) is still somewhat ambiguous. Indeed, $Q$ is always computed for a given principal function $\mathcal{S}$. If we use a different $\widetilde{\mathcal{S}}$ related to $\mathcal{S}$ through a Legendre transformation, we have in general $Q(\mathcal{S})\neq Q(\widetilde{\mathcal{S}})$. We already argued that we expect to find always a factorised form of the $\mathcal{S}$-function, and therefore finding $Q\neq 0$ for a given solution derived from a specific $\mathcal{S}$ does not imply that the solution is not fake supersymmetric: it just means that we are not working in the Legendre frame where the function factorises (up to a constant).


Reference \cite{Trigiante:2012eb} noted that $\mathcal{S}$ can always be shifted with a constant and still result in the same equations of motion for the fields and the metric. Such a constant shift could then be used to set $Q=0$, so the ambiguity in the definition was already present without invoking Legendre transformations. They remarked nevertheless that the constant shift needed to make $Q$ vanish would typically depend on the initial positions for the scalars. This was then argued to be unphysical since one would like to have a single $\mathcal{S}$-function for solutions that only differ in the initial position.\footnote{A flow equation should be interpreted as a first-order differential equation for which the initial position is a boundary condition.} It is our aim here to clarify this and make it precise. The precise definition of restricted fake supersymmetry is the content of the next section.

\subsection{Restricted fake supersymmetry}

We have argued in previous sections that we can always assume the existence of a factorised $\mathcal{S}$-function (\ref{factorisation1}). Let us now consider a flow in the gauge $f=1$:
\begin{equation}\label{boundary}
\dot{\phi}^i =  -h^{ij}\partial_j \mathcal{W}(\phi, \alpha)\,.
\end{equation}
The flow equations can be used as follows: we specify the value of the scalar fields at the initial position (from now on denoted by $\phi_0$), and then the flow equations tell us how the solution flows to the bulk. This implies that the constants $\alpha^a$ should be interpreted as functions of the initial conditions,
\begin{equation}\label{alpha}
\alpha^a = \alpha^a (\phi^i_0, \dot{\phi}^j_0)\,.
\end{equation}
Remark that the indices $i, j$ run from 1 to $n$, whereas the index $a$ runs over maximally $n$ values and depends on how many integration constants were found in solving the HJ equation for $\mathcal{S}$. Our definition of restricted fake susy will not depend on the ability of finding all the independent constants.

The relation (\ref{alpha}) can be deduced from evaluating the flow equations (\ref{boundary}) at the initial position, 
\begin{equation}\label{alpha_2}
\dot{\phi}^i_0 = \dot{\phi}^i_0(\alpha^a,\phi^j_0)\,,
\end{equation}
which implicitly defines (\ref{alpha}). This begs of course the question what this initial position is, since a priori we could choose any point in spacetime. It is natural to restrict to solutions  with an  asymptotic behaviour that correspond to the vacuum state of the theory. Indeed, when searching for black hole solutions or domain wall flows, we are naturally led to require an asymptotic form for the metric, such as flat space for the first and AdS for the latter. We call this asymptotic region from here on `the boundary', and we identify it with what we previously called `the initial position', so that by $\phi^i_0$ we denote the value of the scalar fields at the boundary. As we want the same fake superpotential to describe the whole set of solutions for different values of $\phi^i_0$, we require that the constants $\alpha^a$ which appear in $\mathcal{W}$ do not change as we vary the values of $\phi^i_0$. It turns out that the demand for an asymptotic boundary restricts the dependence of $\alpha^a$ on the initial conditions. We will make this clear after stating the definition of restricted fake supersymmetry:\vspace{10pt}

{\bf General Definition:} \emph{A solution with a specific boundary at infinity is called restricted fake susy (RFS) if the initial positions of scalars that are not restricted by the second-order equations are neither restricted by first-order equations deduced from a principal function in the factorised form $\mathcal{S} = g^{D-1}\mathcal{W}(\phi)$.}\vspace{10pt}

To make this definition more precise we apply or specify it to three different cases: 1) Domain walls with AdS assymptotics (a.k.a. regular domain walls), 2) domain walls with scaling assymptotics and 3) extremal large black holes in asymptotically flat space. In the next sections we provide several illustrations of these definitions and discuss the use of these definitions.\\

\subsubsection{Regular domain walls}

When considering RFS for asymptotically AdS domain walls one has to take into account that the scalar potential fixes certain scalars in the AdS vacuum at the boundary where we fix the geometry (which we will call the UV), and this in turn implies that these scalars start off at a critical point of $\mathcal{W}$.\footnote{Take for instance the gauge $f=g^{D-1}$. Then $f\rightarrow \infty$ near the boundary and the flow equation (\ref{fs1}) implies $\partial_i \mathcal{W}\rightarrow 0$, while the product $f h^{ij}\partial_i\mathcal{W}=\dot{\phi}$ remains finite in the limit.} The moduli-space is parametrized by all other scalars, i.e.~those that are not fixed in the vacuum and span the moduli space that is holographic dual to the space of marginal couplings (a.k.a.~the conformal manifold).  Therefore, the condition that the scalars should be freely chosen at the boundary only holds for the scalars that span the moduli space of the vacuum in the UV. Hence we see that the natural form of the general definition would be:\\

{\bf Definition:} \emph{ A regular domain wall is RFS when it is a solution to a first-order flow generated by a factorised principle function such that the initial position of the flow in the AdS-moduli space in the UV is not fixed.}\\

In order to illustrate this we consider again the solution of a free scalar field in AdS discussed in section \ref{Section-freeAdS}. If we want that solution to be asymptotically AdS with radius $L$, then we must have
\begin{equation}
\left(\frac{\dot{g}}{fg}\right)^2\rightarrow \frac{L^{-2}}{(D-1)(D-2)}
\end{equation}
as we approach the boundary. The equation of motion for the metric \eqref{fs2} then implies that
\begin{equation}
\cosh^2\Bigl[\sqrt{\tfrac{D-1}{2(D-2)}}(\phi_0 -\alpha)\Bigr]=1~,
\end{equation}
which leads to $\alpha=\phi_0$. As we see, once we fix $\alpha$ we can no longer choose $\phi_0$ at will and have an AdS asymptotic region at the same time. This solution would still be fake supersymmetric in the original sense, but not RFS. Our definition represents therefore a genuine restriction on the set of fake supersymmetric solutions. Moreover, the flow of the example for general values of $\phi_0$ is singular in the bulk, so we see that the physical requirement of letting the scalars take arbitrary values at the boundary can serve as a way of extracting physical solutions from all the possible fake supersymmetric ones.\\

\subsubsection{Scaling solutions}
A common and physical boundary condition different from AdS is a \emph{scaling solution}. These are solutions conformal to AdS with conformal factors that are a certain power of the radius.\footnote{The cosmological cousins of scaling domain wall solutions are also known as \emph{power law solutions} and are well-studied in the cosmology literature as solutions to exponential potentials.} The geometries are characterized by a real number $p$:
\begin{equation}
\d s^2_4 = \d z^2 + z^{2p}(-\d t^2 + \d \vec{x}^2_{D-2})\,.
\end{equation}
Scaling solutions occur as asymptotic limits of more general solutions, where they are attractors or repellers of the dynamical system. Domain wall scaling solutions (or cosmological scaling solutions) that can be lifted to 10 or 11-dimensional supergravity describe the near horizon of non-conformal Dp-branes (or Sp-branes) \cite{Kanitscheider:2008kd, Trigiante:2012eb}. Scaling solutions therefore play a role in holography as fundamental as AdS space does.  The geometries are called scaling because the scaling symmetry of AdS space is broken, but not entirely. The dilatation symmetry defines a \emph{conformal} Killing vector instead of a normal Killing vector, and is generated by the action
\be z \rightarrow \lambda z\,,\qquad t,x,y, \rightarrow \lambda^{-p}(t,x,y)\,,
\ee
which rescales the metric. Canonically normalised scalar fields then run logarithmically
\be\label{logrun}
\phi^i(z) = a^i\ln(z) + b^i
\ee
for $z\rightarrow 0,\infty$. The numbers $a^i$ and $b^i$ are certain constants. So these scalars become infinite at the boundary and one has to be careful with defining RFS since $\phi^i_0$ becomes ill-defined. The constants $b^i$ can now play the role of the moduli value $\phi_0$ since this constant determines the value of the scalars at the reference value $z=1$,\footnote{This reference value is set at $z=1$ because we set the boundary at $z=0$. In general, we could fix the boundary at $z=z_0$ by writing $\phi^i(z) = a^i\ln(z-z_0) + b^i$, with which the $b^i$'s would become the value of the scalars at $z=1+z_0$.} and the second-order equations do not fix this constant. The constants $a^i$ on the other hand are fixed by the scaling properties of the solution, and depend on the different parameters that appear in the scalar potential, as was shown for instance in \cite{Copeland:1999cs} (see also \cite{Tolley:2007nq,Chemissany:2007fg}). Hence the definition of RFS for solutions that are asymptotically scaling is that $b^i$ should be free to choose and not fixed by the superpotential.  The reason behind this definition is as before: if the fake susy equations are really to be used as flow equations, then $b$ should be the integration constant of the first-order flow and should not be fixed.

The discussion above treats canonically normalized scalars. The definition for more general flows in which not all scalars are canonically normalized captures the same idea and goes as follows:

{\bf Definition:}  \emph{A domain wall solution that is asymptotically scaling is called RFS when it is a solution to a first-order flow generated by a factorised principal function such that the value of the scalars at the reference position for $z$ is not fixed by the flow.}\\

\subsubsection{Extremal large black holes}
We consider theories in 4D of gravity coupled to Abelian vectors and massless scalars. Such theories have spherically symmetric black hole solutions at zero temperature, which share many similarities with domain walls (at zero temperature) in that there exist analogous first-order flow equations in terms of a fake superpotential $\mathcal{W}$ when Hamilton's principal function factorises as we review in Appendix \ref{black holes}. 

The metric Ansatz for spherically symmetric black holes at $T=0$ in $D=3+1$ is:
\be\label{blackholeansatz}
\d s_4^2 = -e^{2U(\tau)}\d t^2  + e^{-2U(\tau)}[\frac{\d \tau^2}{\tau^4} + \frac{1}{\tau^2}\d\Omega_2^2]\,.
\ee
The coordinate $\tau$ is the inverse of the familiar radial coordinate $r=\tau^{-1}$. At $\tau=0$ the solution approaches flat space $e^U\rightarrow 1$. ``Large" extremal black holes have regular horizons at $\tau\rightarrow -\infty$ where the solution becomes AdS$_2\times S^2$. All scalars are massless and there values at the Minkowski boundary are not constrained by the second-order equations of motion. We therefore arrive at the following natural form of the general definition for large black holes:\\

{\bf Definition:}  \emph{An extremal large black hole is called RFS when it is a solution to a first-order flow generated by a factorised principle function such that the values of all scalars at spatial infinity are not fixed by the flow.}\\

For large extremal black holes the scalars flow to fixed horizon values that are local critical points of the black hole potential $V_{BH}$:
\begin{equation}\label{critical}
\dot{\phi}^i_{H}= 0 \,,\qquad \partial_i V|_{H}=0\,.
\end{equation}
The values of the scalars at the horizon are also at a critical point of $\mathcal{W}(\phi;\alpha)$,
\begin{equation}\label{critical2}
\partial_{\phi^i} \mathcal{W}(\phi;\alpha)|_{H} =0 \,.
\end{equation}
Since $V$ does not depend on $\phi^i_0$, neither will the value of the scalars at the horizon. Therefore equation (\ref{critical2}) must be such that $\alpha$ does not depend in an essential way on $\phi^i_0$ since otherwise the value of the scalars at the critical point would depend on $\phi^i_0$. Hence we reproduce the condition of RFS purely from the point of view of the attractor mechanism that is required for regular black holes. This was already hinted upon in \cite{Trigiante:2012eb}, using the restriction of fake  supersymmetry based on computing $Q$.

We remark, however, that even though attractor flows must be RFS, the converse does not generally hold. In Appendix \ref{black holes} we describe a solution that is RFS and at the same time has a naked singularity in the bulk without a clear physical interpretation. So RFS is not a sufficient condition for solutions being physical, it is however a necessary condition. 

To our knowledge the explicit fake superpotentials that have been constructed in the literature for attractor extremal black holes (e.g.~\cite{Ceresole:2007wx, Andrianopoli:2007gt, Bossard:2009we}) are of course RFS and do not have any free parameter $\alpha$. A priori it is not inconsistent to have a free parameter $\alpha$ in $\mathcal{W}$ despite the fact that the extrema of $\mathcal{W}$ can only depend on magnetic ($p^I$) and electric ($q_I$) charges, since near the attractor values $\phi_H$, $\alpha$ could appear in $\mathcal{W}$ for instance as
\be
\mathcal{W} = \alpha (\phi-\phi_H)^N + \ldots
\ee
with $N>1$. However since not a single example is known we believe that a free $\alpha$ for attractor black holes is excluded.\footnote{One could perhaps even argue that such an $\alpha$ would violate the no-hair theorem.}

\section{Examples with scaling asymptotics}\label{ex}

We present examples which have scaling assymptotics.  The first one concerns the axion-dilaton scaling solution from \cite{Sonner:2007cp}. We show how it can be derived from a fake superpotential, and we give a new analytic solution which interestingly interpolates between two scaling regimes. The second example treats the domain walls obtained from Abelian gaugings which were found in \cite{Blaback:2013taa}. Also here we have been able to find a new analytic solution, which we analyse in the light of RFS.  In the appendix we discuss a domain wall solution already constructed in \cite{Kuperstein:2014zda}.

\subsection{The axion-dilaton model} \label{sectionSTmodel}
We now consider a model by Sonner and Townsend \cite{Sonner:2007cp} with two real scalars subject to an exponential potential  
\begin{equation}
S=\int(\star1)\left[\mathcal{R}-\tfrac{1}{2}(\partial \phi)^2-\tfrac{1}{2}e^{\mu\phi}(\partial \sigma)^2-\tfrac{1}{2}\Lambda e^{\lambda\phi}\right].
\end{equation}
Here, $\phi$ and $\sigma$ are respectively the dilaton and the axion, and $\lambda$ and $\mu$ are constants. Following the notation of \cite{Trigiante:2012eb}, we work in $D=4$, use the gauge $f=g^3$ and furthermore denote $g(z)=\exp(\tfrac{1}{\sqrt{3}}U(z))$. There are two scaling solutions. The simplest scaling solution has a constant axion:
\begin{align}
 U(z)= -\frac{\sqrt3}{2} \log z +U_0\,,\quad \phi(z)=\frac{3\Lambda -2}{\lambda} \log z+\phi_0\,,
\end{align}
where $U_0,\phi_0$ are constants that obey
$ \lambda^2\Lambda\exp(2\sqrt{3}U_0 +\lambda\phi_0) = 3\Lambda -2$. 

The solution with a non-constant axion is given by:
\begin{align}
 U(z)=a\log z +U_0\,,\quad \phi(z)=\frac{2}{\mu}\log z+\phi_0\,,\quad 
 \sigma(z)=-\frac{d}{z}-\sigma_0\,,
\end{align}
where the various constants are related as follows
\begin{align}
&a=-\frac{1}{\sqrt{3}}\left(1+\frac{\lambda}{\mu}\right)\,,\nonumber\\ & d^2=4e^{-\mu\phi_0}\left[-\frac{1}{\mu^2}+\frac{\lambda}{3\mu}\left(1+\frac{\lambda}{\mu}\right)\right]\,,\nonumber  \\
& 4\left(1+\frac{\lambda}{\mu}\right)=- 3\Lambda e^{2\sqrt{3}U_0+\lambda\phi_0}~. \label{relationScaling2}
\end{align}

The first scaling solution can easily be shown to be RFS so we focus the discussion on the second scaling solution. Despite the simplicity of the action, a fake superpotential for the second scaling solution has not been found in the literature. We were able to find explicit fake superpotentials for certain combinations of the constants $\lambda$ and $\mu$, namely $\lambda+\mu=\pm\sqrt{3}$ and $\lambda+2\mu=\pm\sqrt{3}$.\footnote{To our knowledge these values do not have any special physical meaning.} We will focus our discussion on the case $\lambda+\mu=-\sqrt{3}$, for which we found
\begin{equation}
\mathcal{W}(\phi,\sigma,\alpha)=-\sqrt{\Lambda}e^{-\frac{\sqrt{3}}{2}\phi}(\sigma+\alpha)~. \label{eq:AxioDila}
\end{equation}
The scaling solution then corresponds to a solution of the flow equations only if $\alpha=\sigma_0$. This solution is therefore \emph{not} RFS. 

Once we have the fake superpotential which provides us with first-order equations of motion, the next logical step is to try to find a more general analytic solution for the model. We have succeeded in finding a solution that generalises the scaling one:
\begin{align}
\sigma(z)&=-\alpha+A\tan\left[\frac{\sqrt{3}\mu AC}{8}(z-B)\right]\,, \nonumber\\
\phi(z)&=-\frac{1}{\mu}\log\left[\frac{\sqrt{3}\mu A^2}{4}\right]-\frac{2}{\mu}\log\sec\left[\frac{\sqrt{3}\mu AC}{8}(z-B)\right]\,, \nonumber\\
U(z)&=\frac{1}{\sqrt{3}}\log\frac{C}{2\sqrt{\Lambda}}+\frac{1}{2}\phi(z)~, \label{generalSolution1}
\end{align}
with $A,B$ and $C$ integration constants.

It is interesting to note that this more general solution \eqref{generalSolution1} interpolates between two scaling solutions. To see this, let for instance $B=\pi/2$ and $C=8/\sqrt{3}\mu A$. Then for small $0<z\ll1$ we find
\begin{align}
\sigma(z)&\simeq-\alpha-\frac{A}{z}\,,\nonumber \\
\phi(z)&\simeq \frac{2}{\mu}\log z-\frac{1}{\mu}\log\left[\frac{\sqrt{3}\mu A^2}{4}\right].
\end{align}
In order to recover the scaling solution, we see that we should let $\alpha=\sigma_0$ and\footnote{Since the relations \eqref{relationScaling2} only fix $d$ up to a minus sign, a scaling solution can have both $\pm d$ in it.} $A=\pm d$. For definiteness we take $A=d$, so that $A^2=4e^{-\mu\phi_0}/\mu\sqrt{3}$ and then
\begin{align}
\sigma(z)\simeq-\sigma_0-\frac{d}{z} \,,\quad \phi(z)\simeq \frac{2}{\mu}\log z+\phi_0 \,,\quad
U(z)\simeq\frac{1}{\mu}\log z +U_0\,,
\end{align}
which reproduces the scaling solution. 

\begin{figure}
\includegraphics[scale=0.9]{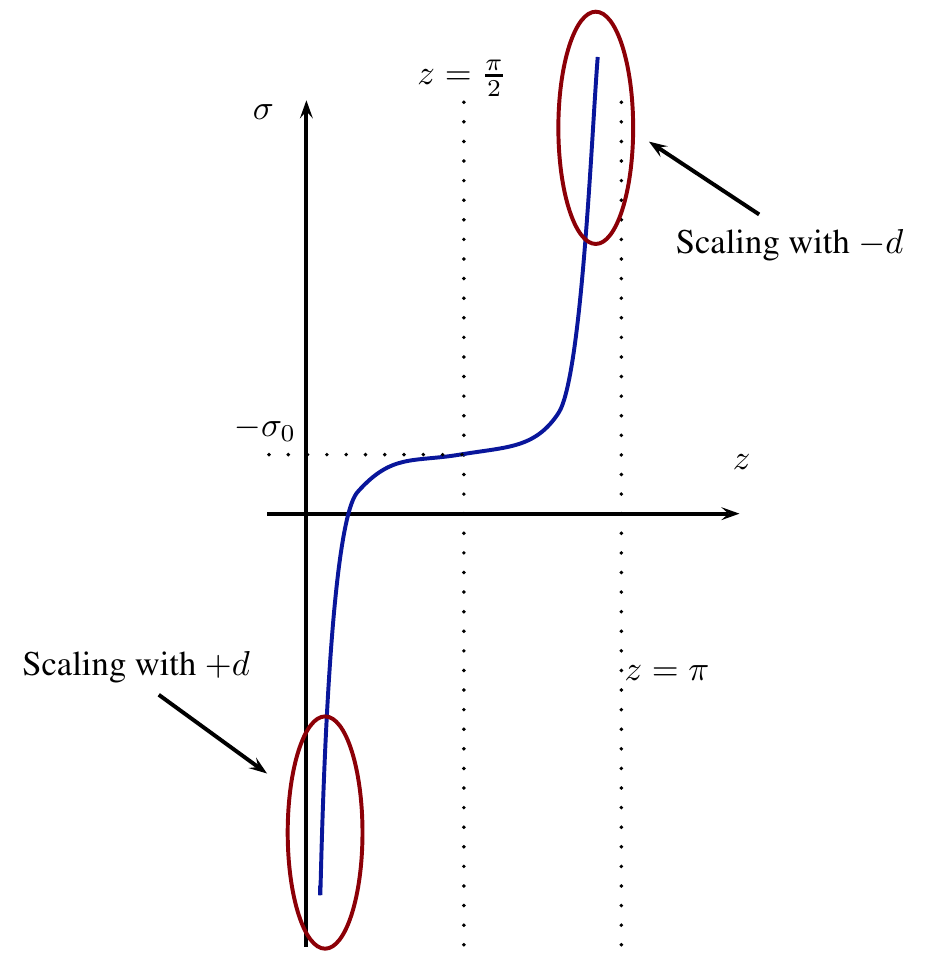} 
\includegraphics[scale=0.9]{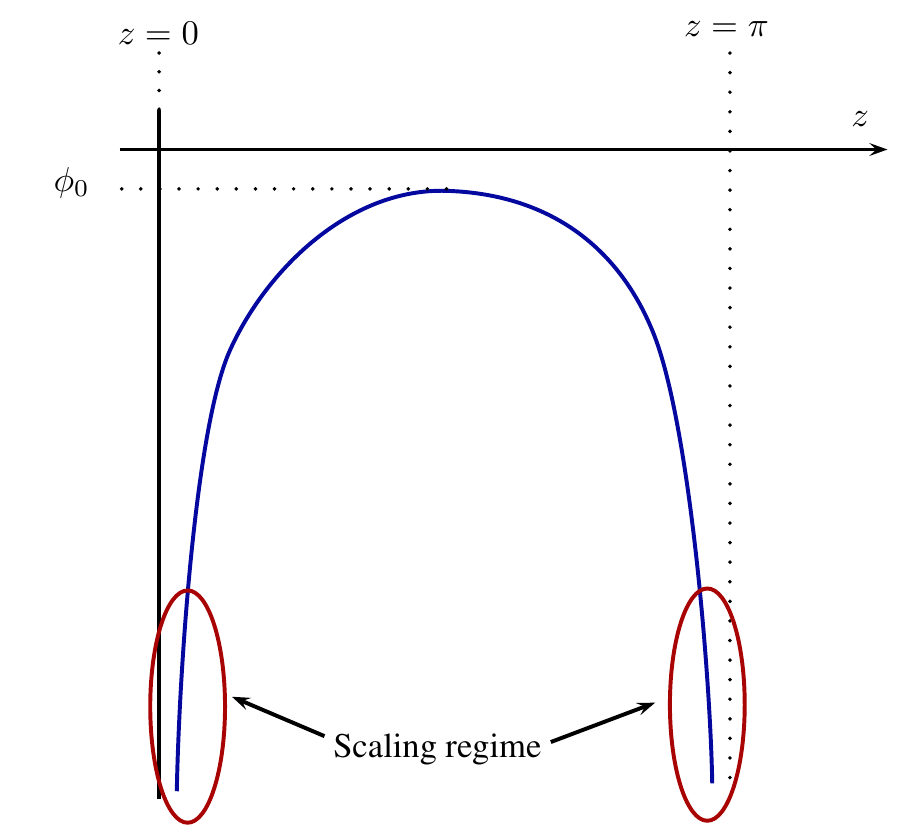}
\caption{The interpolating flow described by \eqref{generalSolution1}.}\label{fig-interpolation}
\end{figure}

Now we let $z$ flow towards $z\simeq\pi$:
\begin{align}
\sigma(z')\simeq-\sigma_0+\frac{d}{z'} \,,\quad
\phi(z')\simeq \frac{2}{\mu}\log z'+\phi_0 \,,\quad
U(z')\simeq\frac{1}{\mu}\log z' +U_0\,,
\end{align}
with $z'=\pi-z$. This is again a scaling solution. As we have just shown, the general solution interpolates between two different scaling regimes, corresponding to the two different values for $d$ allowed by \eqref{relationScaling2}. Figure \ref{fig-interpolation} illustrates this interpolation.

\subsection{Domain walls from Abelian gaugings}\label{Abelian}

We follow section 3 of \cite{Blaback:2013taa} which concerns truncations of Abelian ($SO(2), SO(1,1)$) gauged supergravities in $D=7$.  There are two scalars $u$ and $x$, whose action is
\begin{equation}
S=\int(\star1)\left[\mathcal{R}-\tfrac{1}{2}(\partial u)^2-\tfrac{1}{2}(\partial x)^2-\tfrac{1}{2}\exp\left(-2\sqrt{\tfrac{3}{5}}x\right)\left(|h|e^{-u}-|m|e^u\right)^2\right].
\end{equation}
A fake superpotential was found \cite{Blaback:2013taa} of the form
\begin{equation}
W(x,u,\alpha)=\exp\left(-\sqrt{\tfrac{3}{5}}x\right)\left(\alpha-|h|e^{-u}-|m|e^u\right).
\end{equation}
The case $\alpha=0$ gives supersymmetric solutions \cite{Bergshoeff:2004nq}. We found explicit solutions for general $\alpha$:
\begin{align}
u(z)&=\log\left[\sqrt{\frac{h}{m}}\tanh\left[-\sqrt{hm}(pz+B)\right]\right] \,,\nonumber\\
x(z)&=-\sqrt{\frac{5}{3}}\log\frac{p}{A^6} + \sqrt{\frac{3}{5}}\left[\alpha pz + \log\sinh[-2\sqrt{hm} (p z + B)]\right]\,,\nonumber \\
g(z)& =Ae^{\alpha pz/10}\left(\sinh[-2\sqrt{hm} (p z + B)]\right)^{\frac{1}{10}},
\end{align}
with $A,B$ and $p$ integration constants.  These solutions have not appeared earlier in the literature and they for instance imply the existence of new extremal but non-susy 7-brane solutions of IIB via dimensional oxidation (or new M2 solutions in 11d supergravity) \cite{Bergshoeff:2004nq}.

If we let $B=0$, then the boundary is located at $z=0$. We can expand the solution around the boundary as $z=-\frac{\varepsilon}{2p\sqrt{hm}}$ for small $\varepsilon$, to get
\begin{align}
u(\varepsilon)&=\log\varepsilon + \frac{1}{2}\log\left(\frac{h}{m}\right)-\log 2 + \mathcal{O}(\varepsilon^2) \,,\nonumber \\
x(\varepsilon)&=\sqrt{\frac{3}{5}}\log\varepsilon -\sqrt{\frac{5}{3}}\log\frac{p}{A^6} +\mathcal{O}(\varepsilon) \,,\nonumber \\
g(\varepsilon)& =A\varepsilon^{\frac{1}{10}}+\mathcal{O}\left(\varepsilon\right).
\end{align}
For any value of $\alpha$ we obtain scaling behaviour and  this does not restrict the value of the scalar $x$ at the boundary (the value of $u$ is fixed by the potential).

\section{Implications of RFS}\label{im}

\subsection{Flows with AdS asymptotics}
The definition of RFS has a link with the attractor mechanism: the flows should be independent of the choice of moduli at infinity. It is known (see for instance \cite{Behrndt:2001qa}) that domain walls with AdS asymptotics are attractors and we therefore expect them to be RFS and this is indeed consistent with all models we know from the literature. One can still wonder: are there regular AdS domain walls that have arbitrary constants $\alpha$ in the (fake) superpotential from which they are derived? If the answer to this question is affirmative, we would then have theories with several different domain walls flowing from the same initial AdS vacuum to the same final one. Examples of non-unique domain wall flows between two stable AdS vacua have been found and analysed in \cite{Bobev:2009ms, Bobev:2010ib, Bobev:2011rv, Bobev:2013yra}. In those examples, however, the reason for having multiple flows for the same initial position of the scalars is not due to the existence of a family of fake superpotentials $\mathcal{W}(\phi,\alpha)$. What happens instead is that the metric $h_{ij}$ on the scalar manifold reaches a coordinate singularity in the AdS vacuum in such a way that the flow equation $f^{-1}\dot{\phi}^i = - h^{ij}\partial_j \mathcal{W}$ allows multiple flows
\cite{Bobev:2009ms} derived from the same superpotential. Hence the examples of \cite{Bobev:2009ms, Bobev:2010ib, Bobev:2011rv, Bobev:2013yra} show that the flow between two stable AdS vacua need not be unique. However, the way in which the non-uniqueness is achieved relies on the structure of the scalar manifold rather than on having a family of fake superpotentials.

\subsection{Flows with scaling asymptotics}

The definition of RFS can help to clarify a confusion in the literature concerning scaling solutions. It has been claimed in \cite{Tolley:2007nq} that scaling implies that the scalar fields necessarily follow Killing flows on the scalar manifold $\mathcal{M}$. In other words, there exists a Killing vector field tangent to the curves:
\be \label{Killing}
\nabla_i \dot{\phi}_j + \nabla_j\dot{\phi}_i =0\,.
\ee
Since fake supersymmetry makes the flow a gradient flow (\ref{fs1})
\be
\nabla_i \dot{\phi}_j - \nabla_j\dot{\phi}_i =0\,,
\ee
we find that
\be\label{straightline}
\nabla_i \dot{\phi}_j =0\,,
\ee
and the curves are geodesics \cite{Chemissany:2007fg}.  So fake supersymmetry serves to prove the conjecture that scaling solutions are geodesics \cite{Karthauser:2006ix}. Nevertheless, in \cite{Sonner:2007cp} an axion-dilaton scaling solution was found that is not geodesic, as noted in \cite{Chemissany:2007fg}, and thereby the link scaling-geodesics of \cite{Karthauser:2006ix} seemed to be disproved. It was hence concluded that this is an example of a solution that cannot be fake supersymmetric. This cannot be correct since we already argued that all solutions are fake susy in the standard sense, and indeed in section \ref{sectionSTmodel} we constructed a fake superpotential for the scaling solution of \cite{Sonner:2007cp}. However that solution is not RFS, which explains why there is no Killing flow.

The extra restriction we demand from RFS is what allows the definition of a congruence of curves that correspond to solutions. Indeed, the definition of RFS ensures precisely that \textit{the same} fake superpotential generates all the solutions with different values of $\phi_0$. This is needed in order to define the vector field $\dot{\phi}$ through the congruence of curves generated by the fake superpotential. Hence the reason the flow is neither Killing nor geodesic is clearly related to the absence of RFS. If there is no notion of a tangential vector field $\dot{\phi}$ derived from a fake superpotential $\mathcal{W}$, then the proof in \cite{Tolley:2007nq} that leads to (\ref{Killing}) is not correct. 

Let us verify this explicitly for the superpotential of the axion-dilaton model (\ref{eq:AxioDila}). If we pretend that there is a velocity field then it will be given by $\d\mathcal{W}$. Using the fake superpotential constructed in the previous section one easily verifies that the curve is not Killing nor geodesic:
\begin{equation}
\nabla_\sigma V_\sigma=\partial_\sigma V_\sigma+\Gamma^\phi_{\sigma\sigma}V_\phi=-\frac{\alpha}{(\sigma+p)^2}-\frac{\alpha\mu\sqrt{3}}{4}e^{\mu\phi}\neq 0~,
\end{equation}
where $V$ denotes the tangent vector to the curve.

The other solutions we constructed in section \ref{Abelian} are RFS and it is straightforward to verify that the flow is indeed Killing and geodesic.

Reference \cite{Sonner:2007cp} hinted on a possible link between the consistency of the single scalar truncation of \cite{Celi:2004st} used in section \ref{sec:fact} and fake susy. Although reference \cite{Sonner:2007cp} did not employ our restricted definition of fake supersymmetry, we now show the link is correct, at least for scaling solutions that are RFS. 

Suppose the scaling solution flows indeed along an isometry in scalar field space. Then the metric, after field redefinition for which only one scalar (say $\tilde{\phi}^1$) is active, looks as follows \cite{Tolley:2007nq}:
\be
\d s^2 = \tilde{h}_{11}(\d \tilde{\phi}^1)^2 + 2\sum_{I=2}^N \tilde{h}_{1I}\d \tilde{\phi}^1\d \tilde{\phi}^I + \sum_{I,J=2}^N \tilde{h}_{IJ}\d \tilde{\phi}^I\d \tilde{\phi}^J\,, 
\ee
where $\tilde{h}_{11}, \tilde{h}_{1I}, \tilde{h}_{IJ}$ cannot depend on $\tilde{\phi}^1$. Restricted fake supersymmetry is what allowed us to use the link scaling-Killing and it also implies that the Killing vector dual to $\tilde{\phi}^1$ is hypersurface orthogonal such that the off-diagonal term $\tilde{h}_{1I}$ can be set to zero. Then it can be easily shown \cite{Tolley:2007nq} that the potential must take the form
\be
V = e^{\lambda \tilde{\phi}^1} f(\tilde{\phi}^2,\ldots, \tilde{\phi}^N)\,,
\ee
in order to allow for the scaling solution. In the above action, $\lambda$ is a constant and $f$ is some function of the remaining scalars. The existence of the scaling solution then simply implies that $f(\tilde{\phi}^2,\ldots, \tilde{\phi}^N)$ has a critical point $\partial f=0$, which shows that the one-scalar truncation is consistent since $\tilde{h}_{1I}=0$.

\subsection{Holographic viewpoint}
As already mentioned in the Introduction, the holographic interpretation of scalar profiles along the bulk are RG flows in the dual field theory \cite{DeWolfe:1999cp, deBoer:1999xf, Skenderis:1999mm}. The scalar field $\phi^i$ evaluated at some fixed value of the holographic coordinate $z$ corresponds to a coupling constant $g^i$ in the dual field theory evaluated at an energy scale $\mu$ related to $z$. RG flow equations are then of the form
\be
\frac{\partial g^i}{\partial \ln \mu} = \beta^i (g^1,\ldots g^n)\,.
\ee
Since the right hand side does not contain an explicit $\mu$ dependence it implies that the holographic first-order flow must be such that the scalars decouple from the warpfactor $g$ in the right gauge choice for $f$. Indeed the Hamilton Jacobi equations do exactly that in the gauge $f=1$ when the principal function $\mathcal{S}(\phi,g)$ factorises as in (\ref{factorisation1})\footnote{Although subtleties can arise in this identification between RG flows and beta-functions \cite{Heemskerk:2010hk, Faulkner:2010jy}}. The corresponding superpotential $\mathcal{W}$ should then correspond to Zamolodchikov's C-function \cite{Zamolodchikov:1986gt}, extended to theories in general dimension\footnote{Although it is not yet known whether a C-function exists in general, especially in odd dimensional QFT's.}. Regular domain wall flows between two AdS vacua are dual to quantum field theories that have a UV and IR fixed point.  In even dimensions the C-function is related to anomaly coefficients, which do not depend on the values of marginal couplings since they are topological (see for instance \cite{Duff:1993wm}). This is the statement dual to our notion of restricted fake supersymmetry. 

Our definition of RFS is however not automatically satisfied for flows that have a scaling geometry at infinity and RFS can be a genuine restriction. The quantum field theories dual to such flows are also less understood and it is not certain whether a would-be C-function cannot depend on the value of certain marginal couplings in the UV. It would be interesting to understand whether RFS might be natural from the QFT point of view since that would imply that non-RFS solutions are perhaps non-physical. One obvious set of quantum field theories that are dual to scaling geometries correspond to SYM theory in dimensions different from 4. The holographic description can be found by considering the decoupling limit of D$p$ branes with $p\neq 3$ since the (singular) near horizon geometries are indeed of the scaling type \cite{Kanitscheider:2008kd}. For these QFT's the flows are RFS.

\section{Discussion}\label{discussion}

In this paper we have clarified some confusing aspects about the notion of fake supersymmetry.  Our main result is the observation that fake supersymmetry, as defined in the literature, corresponds to a specific factorisation property of Hamilton's principal function $\mathcal{S}$ (\ref{factorisation1}), that can be shown to always hold  at zero temperature (using for instance Legendre transformations).
This led us to define a restricted version of fake supersymmetry, abbreviated RFS,  that is more than just the local existence of the usual fake superpotential flow equation (\ref{schematic}). RFS invokes the boundary data by insisting that scalar field values which are not fixed by the second-order equations of motion near the boundary should neither be fixed by the first-order equations.

We have argued why the restricted definition is physically relevant by demonstrating its appearance for regular domain walls and attractor black holes. We have suggested that for both domain walls and extremal black holes RFS is necessary for having regular solutions. In the black hole case we know it is not sufficient, for domain walls this could still be and is currently under investigation. In any case this very fact shows that RFS is a useful concept. The main point of our paper is to argue that RFS is a natural restriction that really classifies zero-temperature solutions into two branches: RFS and non-RFS. The standard notion of fake supersymmetry does not do this and is hence not physically meaningful. Our hope is that this observation turns out more useful and has a deeper physical meaning. A natural direction to look into is what the  holographic dual to RFS would be. As we argued it corresponds to a QFT for which the C-function does not depend on the values of the marginal couplings in the UV. This statement seems obvious for RG flows between fixed points but might not be for more general RG flows. If one can demonstrate in QFT that this would be a necessary condition it would imply that non-RFS gravity solutions with a QFT dual are not physical.

Since our restriction relies on the boundary conditions, we have paid special attention to a boundary condition that is not so often discussed in the literature on domain walls: scaling geometries. RFS turns out to explain a confusion in the literature  concerning the connection between the asymptotic motion of the scalars and geodesic flows on the target space.  Contrary to the claims in \cite{Chemissany:2007fg} and \cite{Trigiante:2012eb}, the reason that the scaling solution of \cite{Townsend:2007aw} is not a geodesic on target space is not due to the lack of a factorisable principal function, since that is always possible. We have found that the solution is not a Killing flow, demonstrating a loophole in the proof of \cite{Tolley:2007nq} due to the absence of RFS. If the solution is not a Killing flow then the argument for it being a geodesic flow \cite{Chemissany:2007fg} is not applicable.

Our paper relies strongly on ideas from \cite{Trigiante:2012eb}, which we have made more precise. The crucial ingredient lacking in \cite{Trigiante:2012eb} was the use of Legendre transformations of the function $\mathcal{S}$.
The present work extends \cite{Trigiante:2012eb} also in different ways. For instance, we were able to find the full analytic solution for flat axion-dilaton domain walls of \cite{Sonner:2007cp} that interpolates between scaling geometries (see section \ref{sectionSTmodel}).

\section*{Acknowledgements}
We are happy to acknowledge useful discussions with  T.~Hertog, R.~Monten, S.~Kuperstein and especially N.~Bobev. We thank F.~F.~Gautason, B.~Vercnocke, M.~Trigiante and especially N.~Bobev for comments on an earlier draft. TVR also wishes to thank M.~Trigiante and B.~Vercnocke for an earlier collaboration on this topic. This work is supported by the FWO odysseus grant G.0.E52.14N.  BT is aspirant FWO. We acknowledge support from the European Science Foundation Holograv Network.

\newpage
\appendix

\section{The one-dimensional effective action}\label{effActionApp}
We show here that the action given in \eqref{effAction} reproduces indeed the equations of motion for ansatz derived from the original Einstein-scalar action.
\subsection{Equations of motion for the domain wall ansatz}
The equations of motion which one finds from
\begin{equation}
S = \int\ (\star1) \left[ R - \tfrac{1}{2}h_{ij}(\phi)\partial\phi^i\partial\phi^j - V(\phi) \right]
\end{equation}
are
\begin{align}
R_{\mu\nu}&=\frac{1}{2}h_{ij}\partial_\mu\phi^i\partial_\nu\phi^j+\frac{Vg_{\mu\nu}}{D-2} \\
h_{ij}\square \phi^i&=\partial_j V+\left(\frac{1}{2}\partial_jh_{il}-\partial_lh_{ij}\right)\partial_\mu\phi^i\partial^\mu\phi^l~.
\end{align}
For the domain wall ansatz of \eqref{domainwall} we have
\begin{align}
R_{zz}&=\frac{\dot{f}\dot{k}}{fk}-2\frac{\dot{g}\dot{k}}{gk} +(D-1)\frac{\dot{f}\dot{g}}{fg}-(D-1)\frac{\ddot{g}}{g}-\frac{\ddot{k}}{k}\\
\frac{f^2}{g^2k^2} R_{tt}&=-\frac{\dot{f}\dot{k}}{fk}+D\frac{\dot{g}\dot{k}}{gk}-\frac{\dot{f}\dot{g}}{fg}+\frac{\ddot{g}}{g}+\frac{\ddot{k}}{k}+(D-2)\frac{\dot{g}^2}{g^2}\\
\frac{f^2}{g^2} R_{xx}&=-\frac{\dot{g}\dot{k}}{gk}+\frac{\dot{f}\dot{g}}{fg}-\frac{\ddot{g}}{g}-(D-2)\frac{\dot{g}^2}{g^2}~,
\end{align}
while all the other components of the Ricci tensor vanish. We are therefore left with three independent equations of motion coming from the Einstein equations. We will take the following linear combinations:
\begin{align}
\frac{f^2}{g^2k^2} R_{tt}+\frac{f^2}{g^2} R_{xx}&=-\frac{\dot{f}\dot{k}}{fk}+(D-1)\frac{\dot{g}\dot{k}}{gk}+\frac{\ddot{k}}{k}=0 \\
R_{zz}+\frac{f^2}{g^2k^2}R_{tt}&=(D-2)\left(\frac{\dot{g}\dot{k}}{gk}+\frac{\dot{f}\dot{g}}{fg}-\frac{\ddot{g}}{g}+\frac{\dot{g}^2}{g^2}\right)=\frac{1}{2}h_{ij}\dot{\phi}^i\dot{\phi}^j ~,
\end{align}
and keep the equation
\begin{equation}
\frac{f^2}{g^2} R_{xx}=-\frac{\dot{g}\dot{k}}{gk}+\frac{\dot{f}\dot{g}}{fg}-\frac{\ddot{g}}{g}-(D-2)\frac{\dot{g}^2}{g^2}=\frac{Vf^2}{D-2}~.
\end{equation}
The equations of motion from the original action for the domain-wall ansatz are hence
\begin{align}
&-\frac{\dot{f}\dot{k}}{fk}+(D-1)\frac{\dot{g}\dot{k}}{gk}+\frac{\ddot{k}}{k}=0 \label{einstein1} \\
&(D-2)\left(\frac{\dot{g}\dot{k}}{gk}+\frac{\dot{f}\dot{g}}{fg}-\frac{\ddot{g}}{g}+\frac{\dot{g}^2}{g^2}\right)=\frac{1}{2}h_{ij}\dot{\phi}^i\dot{\phi}^j \label{einstein2}\\
&-\frac{\dot{g}\dot{k}}{gk}+\frac{\dot{f}\dot{g}}{fg}-\frac{\ddot{g}}{g}-(D-2)\frac{\dot{g}^2}{g^2}=\frac{Vf^2}{D-2}\label{einstein3}
\end{align}
together with the scalar equation of motion
\begin{equation}\label{eqScalar}
\frac{h_{ij}}{g^{D-1}kf}\frac{\d}{\d z}\left(\frac{g^{D-1}k}{f}\dot{\phi}^i\right)=\partial_j V+\frac{1}{f^2}\left(\frac{1}{2}\partial_jh_{il}-\partial_lh_{ij}\right)\dot{\phi}^i\dot{\phi}^l~.
\end{equation}

\subsection{The effective action}
We will now show that the effective action
\begin{equation}
I[G,k,f,\phi]\int\d z\left\{\frac{G}{f}\left[\frac{D-2}{D-1}\left(\frac{\dot{G}^2}{G^2}-\frac{\dot{k}^2}{k^2}\right)-\frac{1}{2}h_{ij}\dot{\phi}^i\dot{\phi}^j\right]-GfV \right\}
\end{equation}
leads to the same equations of motion after substituting $G=g^{D-1}k$. The Euler-Lagrange equation for $k$ is
\begin{equation}
\frac{G\dot{k}}{fk^3}+\frac{\d}{\d z}\left(\frac{G\dot{k}}{fk^2}\right)=0~.
\end{equation}
We get then
\begin{align}\label{equivEq}
0=\frac{g^{D-1}\dot{k}}{fk^2}+\frac{\d}{\d z}\left(\frac{g^{D-1}\dot{k}}{fk}\right)=\frac{1}{k}\frac{\d}{\d z}\left(\frac{g^{D-1}\dot{k}}{f}\right)~,
\end{align}
which is equivalent to \eqref{einstein1}.

The Euler-Lagrange equation for $G$ is
\begin{equation}
\frac{2f(D-2)}{D-1}\frac{\d}{\d z}\left(\frac{\dot{G}}{fG}\right)+\frac{D-2}{D-1}\left(\frac{\dot{G}^2}{G^2}+\frac{\dot{k}^2}{k^2}\right)=-\frac{1}{2}h_{ij}\dot{\phi}^i\dot{\phi}^j-f^2V~,
\end{equation}
while for $f$ we find
\begin{equation}
\frac{D-2}{D-1}\left(\frac{\dot{G}^2}{G^2}-\frac{\dot{k}^2}{k^2}\right)=\frac{1}{2}h_{ij}\dot{\phi}^i\dot{\phi}^j-f^2V~.
\end{equation}
We can combine these two equations to get
\begin{align}
f\frac{\d}{\d z}\left(\frac{\dot{G}}{fG}\right)+\frac{\dot{G}^2}{G^2}&=-\left(\frac{D-1}{D-2}\right)f^2V \label{effAcPot} \\
f\frac{\d}{\d z}\left(\frac{\dot{G}}{fG}\right)+\frac{\dot{k}^2}{k^2}&=-\left(\frac{D-1}{D-2}\right)\frac{1}{2}h_{ij}\dot{\phi}^i\dot{\phi}^j~.
\end{align}
If we remark that
\begin{align}
\frac{\dot{G}}{G}&=(D-1)\frac{\dot{g}}{g}+\frac{\dot{k}}{k}\\
f\frac{\d}{\d z}\left(\frac{\dot{G}}{fG}\right)&=(D-1)\frac{\d}{\d z}\left(\frac{\dot{g}}{g}\right)+\frac{\ddot{k}}{k}-\frac{\dot{k}^2}{k^2}-(D-1)\frac{\dot{f}}{f}\frac{\dot{g}}{g}-\frac{\dot{f}}{f}\frac{\dot{k}}{k} \\
&=(D-1)\frac{\d}{\d z}\left(\frac{\dot{g}}{g}\right)-\frac{\dot{k}^2}{k^2}-(D-1)\frac{\dot{f}}{f}\frac{\dot{g}}{g}-(D-1)\frac{\dot{g}}{g}\frac{\dot{k}}{k}~,
\end{align}
we see that the second equation of motion becomes
\begin{equation}\label{tussenstapEffA}
(D-1)\frac{\d}{\d z}\left(\frac{\dot{g}}{g}\right)-(D-1)\frac{\dot{f}}{f}\frac{\dot{g}}{g}-(D-1)\frac{\dot{g}}{g}\frac{\dot{k}}{k}=-\left(\frac{D-1}{D-2}\right)\frac{1}{2}h_{ij}\dot{\phi}^i\dot{\phi}^j~,
\end{equation}
which can also be written as
\begin{equation}
\frac{1}{2}h_{ij}\dot{\phi}^i\dot{\phi}^j=(D-2)\left[\frac{\dot{g}}{g}\left(\frac{\dot{k}}{k}+\frac{\dot{f}}{f}\right)-\frac{\d}{\d z}\left(\frac{\dot{g}}{g}\right)\right]~.
\end{equation}
This is nothing else than \eqref{einstein2}. In getting \eqref{tussenstapEffA} we used \eqref{equivEq}. A little bit more work reveals that \eqref{effAcPot} is equivalent to \eqref{einstein3}.

Only the scalar equations of motion rest to be checked. Euler-Lagrange for $\phi^j$ gives
\begin{equation}
\frac{\d}{\d z}\left(\frac{G}{f}h_{ij}\dot{\phi}^i\right)=Gf\partial_j V+\frac{G}{2f}(\partial_j h_{il})\dot{\phi}^i\dot{\phi}^l~.
\end{equation}
The left-hand side is
\begin{align}
\frac{\d}{\d z}\left(\frac{g^{D-1}k}{f}h_{ij}\dot{\phi}^i\right)=h_{ij}\frac{\d}{\d z}\left(\frac{g^{D-1}k}{f}\dot{\phi}^i\right)+\frac{g^{D-1}k}{f}(\partial_lh_{ij})\dot{\phi}^i\dot{\phi}^l~.
\end{align}
Therefore we are left with
\begin{equation}
\frac{h_{ij}}{g^{D-1}kf}\frac{\d}{\d z}\left(\frac{g^{D-1}k}{f}\dot{\phi}^i\right)=\partial_j V+\frac{1}{f^2}\left(\frac{1}{2}\partial_jh_{il}-\partial_lh_{ij}\right)\dot{\phi}^i\dot{\phi}^l~,
\end{equation}
which is precisely \eqref{eqScalar}.

\section{Black holes in massless theories}\label{black holes}
In this section we discuss RFS for black hole solutions. We start by reviewing the set-up which allows us to use Hamilton-Jacobi theory in a similar way as for domain walls. 

\subsection{Set-up} 
We consider black hole solutions of theories in 4D featuring Abelian vector fields and massless scalar fields. After integrating out the vector fields in terms of the electric and magnetic charges, one obtains an effective one-dimensional action for the scalars $\phi^i$ and the warpfactor $U$:
\be
I[U,\phi] = \int \d\tau \, \Bigl(4\dot{U}^2 + h_{ij}\dot{\phi}^i\dot{\phi}^j - e^{2U}V(\phi)\Bigr)\,,
\ee
supplemented with the Hamiltonian constraint $4\dot{U}^2 + h_{ij}\dot{\phi}^i\dot{\phi}^j + e^{2U}V(\phi) = 0$.
When Hamilton's principal function factorises as $S(U,\phi)=e^U\mathcal{W}(\phi)$ (see \cite{Trigiante:2012eb}), the Hamilton-Jacobi equations become the following first-order flow equations 
\begin{align}\label{bhflow1}
\dot{\phi}^i =  e^{U}h^{ij}\partial_j \mathcal{W}\,,\qquad 4\dot{U} = e^U \mathcal{W}~,
\end{align}
where the fake superpotential obeys
\be\label{bhflow2}
-V(\phi) = (\partial \mathcal{W})^2 + \tfrac{1}{4}\mathcal{W}^2\,.
\ee

The asymptotic region ($\tau=0$) is Minkowski, such that $e^U\rightarrow 1$ for $\tau=0$. This condition does not constrain $\dot{\phi}^i$ nor $\dot{U}$ at the boundary in the sense that they can take any value consistent with the zero energy condition.  The scalars flow towards the black hole horizon under the influence of the black hole potential, generated by the vector fields. If the extremal black hole is \textit{large}, the horizon is a smooth AdS$_2\times S^2$ solution where the values of the scalars, denoted by $\phi^i_H$, are fixed by the attractor mechanism (up to certain moduli) in function of the electromagnetic charges $(p^I, q_I)$:
\be
\phi^i_H= \phi_H^i(p^I, q_I)\,.
\ee
If the extremal black hole is \textit{small}, the horizon has zero size and coincides with the singularity, which becomes lightlike. Then the scalars still flow at the horizon, but the way in which they flow is highly restricted: the black hole solution becomes \emph{scaling} at the singular horizon \cite{Trigiante:2012eb}. Scaling here means that the horizon geometry is conformal to AdS$_2\times S^2$ with a power-law conformal factor:
\be
\d s^2 =  r^{\alpha}L^2\Bigl(-r^2\d t^2+  \frac{(\d r)^2}{r^2} + \d\Omega_2^2 \Bigr)~.
\ee
Just as for domain walls, we characterize RFS as the property that the values of the scalars at the boundary (not the horizon) can be chosen at will for both large and small black holes. We present now an example which can help in clarifying this discussion.

\subsection{Free scalar in Einstein-Maxwell}

RFS does not seem to be a restriction when applied in the black hole context. 
Take for instance Einstein-Maxwell theory with a decoupled massless scalar field
\be
S =\int (\star1)\left[\mathcal{R}-\tfrac{1}{4} F^2 -\tfrac{1}{2}(\partial\phi)^2 \right]\,.
\ee
The black hole effective potential is constant $V = -q^2$ and can be derived from the following fake superpotential
\be
\mathcal{W}(\phi)=  2q\sin(\tfrac{1}{2}\phi-\alpha)\,.
\ee
With this, we can find solutions that are  asymptotically Minkowski. In terms of the standard radial coordinate $r=\tfrac{1}{\tau}$, the metric is
\begin{align}
\d s^2&=-f(r)\d t^2+f(r)^{-1}\left[\d r^2+r^2\d \Omega_2^2\right]~, \\
f(r)&=\frac{\cos^2(b-\alpha)}{\cos^2\left[(b-\alpha)+\tfrac{q}{2}\cos(b-\alpha)\frac{1}{r}\right]}~,
\end{align}
while the scalar is simply
\begin{equation}
\phi(r)=2b+\frac{q\cos(b-\alpha)}{r}~.
\end{equation}
The parameter $b$ is an integration constant. For $r\rightarrow\infty$ the metric becomes flat space and the scalar flows to a constant $\phi\rightarrow 2b$. This means that for any $\alpha$, we can freely choose the value of the scalar at infinity and still have an asymptotically flat metric, so that this solution is indeed RFS. But it certainly is not a black hole because of a naked singularity in the bulk.

Interestingly, for the special case $2b=2\alpha+\pi$, the flow becomes
\begin{align}
\phi(R)&=2\alpha+\pi~,\\
\d s^2&=-\left(1-\frac{q}{2R}\right)^2\d t^2+\left(1-\frac{q}{2R}\right)^{-2}\d R^2+R^2\d \Omega_2^2~,
\end{align}
where we shifted the $r$ coordinate as $R=r+\tfrac{q}{2}$. This is nothing else than the Reissner-Nordstr\"om solution with a constant scalar field.

\section{Reduced fractional branes}\label{frac}
This example is taken from \cite{Kuperstein:2014zda} (for $p=3$) and we adopt the notation of that reference. Consider gravity coupled to three scalars $\phi, \varphi$ and $b$ in $D=4+1$ dimensions:
\begin{equation} \label{action}
I  = \int (\star1)\left[ R_{5} -\tfrac{1}{2}(\partial\varphi)^2 -\tfrac{1}{2}(\partial\phi)^2 - \tfrac{1}{2}e^{\frac{12 a}{5}\varphi -\phi}(\partial b )^2 - V(\varphi,\phi, b)\right] \,,
\end{equation} 
where $a=\frac{1}{4}\sqrt{\frac{5}{3}}$ and the scalar potential is given by 
\begin{equation}
V(\varphi,\phi, b)  =  -20e^{\tfrac{16 a}{5}\varphi} + \tfrac{1}{2}m^2 e^{\tfrac{28 a}{5}\varphi +\phi}  + \tfrac{1}{2}(Q+mb)^2 e^{8a\varphi}\,.\label{potential}
\end{equation}
The parameter $m$ has the interpretation of a flux quantum and $Q$ of a brane charge.  
The known ``BPS-like'' flat domain walls of these theories lift to so-named fractional D$3$ solutions \cite{Herzog:2000rz} in $10d$. The ``BPS'' flow for those solutions can be obtained from the following principal function 
\begin{equation} \label{S0}
\mathcal{S}_0(g, \phi, \varphi, b) = g^{4}\Bigl((Q+mb)\,e^{4a\varphi} \pm 10 e^{\tfrac{8}{5} a\varphi}\Bigr)\,.  
\end{equation} 
This principal function indeed factorises (\ref{factorisation1}) such that the term between brackets defines the superpotential $\mathcal{W}(\phi,\varphi, b)$ (\ref{factorisation1}).

In \cite{Kuperstein:2014zda} a second \emph{family} of factorisable solutions to the Hamilton-Jacobi equation was found:
\begin{equation} 
\label{S1}
\mathcal{S}_1(\alpha) = g^{4} \left(\, (Q+mb) \, \,e^{4 a\varphi}  \pm 10 \, e^{\tfrac{8}{5} a\varphi} \cosh \left( \sqrt{\tfrac{2}{5}} \phi + \alpha \right) \right)\,,
\end{equation}
where $\alpha$ is an arbitrary real constant. The principal function $\mathcal{S}_1(\alpha)$ does not obviously describe solutions corresponding to one-parameter deformations of the BPS solutions since there is no clear limit for the integration constant $\alpha$ for which $\mathcal{S}_1(\alpha)$ becomes $\mathcal{S}_0$. By Legendre transforming $\mathcal{S}_1(\alpha)$ we can generate an equivalent family of principal functions $\mathcal{S}_2(\beta)$,:
\begin{align} \label{S2}
\mathcal{S}_2(\beta) &=\,(Q+mb) \,g^{4} \,e^{ 4a\varphi} \pm 2\sqrt{\beta^2+25 g^{8}e^{\tfrac{16}{5}a \varphi}} +2\beta\left[\sqrt{\tfrac{2}{5}}\phi+ \tfrac{8a}{5}\varphi + 4\ln(g) \right]\nonumber \\ & - 2\beta\ln\left(  \sqrt{\beta^2+15 g^{8}e^{\tfrac{16}{5}a \varphi}} \pm \beta\right) \, ,
\end{align} 
When $\beta=0$ we do recover $\mathcal{S}_0$. Whereas $\mathcal{S}_1(\alpha)$ is of the factorised form, $\mathcal{S}_2(\beta)$ is not. To verify whether the fractional brane solutions with $\alpha$ different from zero are RFS we need to impose a boundary condition. If we insist on the same boundary geometry as the extremal fractional branes of \cite{Herzog:2000rz} (which for instance is necessary for a holographic interpretation) it was shown in \cite{Kuperstein:2014zda} that this fixes
\begin{equation}
\alpha =  -\sqrt{\frac{2}{5}}\phi_0\,.
\end{equation}
Hence the solution is not RFS. The second principal function does not fix $\beta$ entirely by $\phi_0$, but then the principal function does not factorise. To date it is not clear whether this one-parameter family of solutions are physical, but at least their singularities are identical to those of the extremal fractional brane solutions, such as the Klebanov-Tseytlin (KT) solution \cite{Klebanov:2000nc}, and an interesting attempt to understand them as holographic backgrounds can be found in \cite{Bertolini:2015hua}.

The KT solution is one of the simplest examples of holographic backgrounds without AdS asymptotics. One can readily verify that the reduction of the KT solution (and its deformations discussed here) to a domain wall in 5d has an asymptotic geometry that is scaling, which is another indication of how universal and relevant scaling solutions are. We present the details of the scaling solution for completeness:\footnote{This is for the KT solution with a non-zero constant term in the ``harmonic''.}
\begin{align}
& g \sim z^{5/4}\,,\qquad \e^{2\beta\varphi} \sim r^2\,,\nonumber \\
& \phi = \text{constant}\,,\qquad  b = b_0 + mg_s \ln z \,.
\end{align}
In this gauge $f= z^{5/3}$ and $z$ corresponds then to the standard radial coordinate in 10d. One can verify that also in the gauge $f=1$ the warpfactor $g$ is a power-law.

\small{
\providecommand{\href}[2]{#2}\begingroup\raggedright\endgroup}
\end{document}